\documentclass[pre,twocolumn,groupedaddress,floatfix]{revtex4-1}
\usepackage{amssymb,amsmath}
\usepackage{graphicx}
\usepackage{subfigure}
\usepackage[english]{babel} 
\usepackage{float}
\usepackage{color}

\usepackage{lipsum}
\begin{document}
\newcommand{\be}{\begin{equation}}
\newcommand{\ee}{\end{equation}}
\newcommand{\rojo}[1]{\textcolor{red}{#1}}

\title{The fractional nonlinear impurity: A Green function approach}

\author{Mario I. Molina}
\affiliation{Departamento de F\'{\i}sica, Facultad de Ciencias, Universidad de Chile, Casilla 653, Santiago, Chile}

\date{\today }

\begin{abstract} 
We use a lattice Green function approach to study the stationary modes of a linear/nonlinear (Kerr) impurity embedded in a periodic one-dimensional lattice where we  replace the standard discrete Laplacian by a fractional one. The energies and the mode profiles are computed in closed form, for different fractional exponents and different impurity strengths. The energies of the impurity mode lie outside the linear band whose bandwidth decreases steadily as the fractional exponent decreases. For any fractional exponent values, there is always a single bound state for the linear impurity while for the nonlinear (Kerr) case, up to two bound states are possible, for impurity strengths above certain threshold. 
The energy of the linear mode (or that of the upper energy nonlinear one), becomes directly proportional to the impurity strength at large impurity strengths. 
The transmission of plane waves is also computed in closed form for several fractional exponents, and various impurity strengths. We observe that fractionality tends to increase the overall transmission. The selftrapping transition for the nonlinear impurity shifts to lower nonlinearity values as the fractional exponent is decreased. In both cases, linear and nonlinear, we observe a form of trapping at zero impurity strength, which can be  explained by the near-degeneracy of the spectrum in the limit of a small fractional exponent.

\end{abstract}

\maketitle

{\bf 1.\ Introduction}.\\ 

One of the most basic aspects of a linear discrete and periodic system, such as a chain of atoms or an optical waveguide array, is that its eigenvalues form a well-defined band while their eigenvectors are extended waves labeled unambiguously by their  wavenumbers\cite{kittel}. When a linear impurity is added to the system, the breaking of the (discrete) translational invariance causes one of the states at the band edge to detach from the band giving rise to a localized mode centered at the impurity position. It has been proven that for 1D and 2D lattices, there is always a localized bound state centered at the impurity\cite{slater,harrison}, no matter how weak the strength of the impurity.  The rest of the modes remain extended although they are no longer sinusoidal. The study of a single impurity is usually the first step towards understanding more complex, disordered systems, such as the case when there is a finite fraction of impurities in a system, where the main phenomenon of study is Anderson localization\cite{economou1,economou2}. Some examples of linear impurities include coupling defects, junction defects between two optical or network arrays\cite{miro}, discrete networks for routing and switching of discrete optical solitons\cite{christo}, and also in simple models for magnetic metamaterials, modeled as periodic arrays of split-ring resonators, where magnetic energy can be trapped at impurity positions\cite{wang}.

When nonlinearity is added to a periodic system, mode localization and self-trapping of energy can occur. This localized mode which exists in this completely periodic but nonlinear system is known as a discrete soliton. This concentration of energy on a small region increases with the nonlinearity strength and, as a consequence, the nonlinear mode becomes effectively decoupled from the rest of the lattice. In this high nonlinearity limit, the effect nonlinearity is concentrated in a small region around the soliton, and thus, we could consider the rest of the lattice as approximately linear. We are then left with a linear lattice containing a nonlinear impurity. This simplified system is more amenable to theoretical analysis where closed-form solutions are sometimes possible. In  condensed matter, these nonlinear impurities appear as the result of doping of materials with atoms or molecules that have strong local couplings. In an optical context, the system of interest is a dielectric waveguide array, where one of the guides is judiciously doped with an element with strong polarizability. A more recent example is magnetic metamaterials, where the  system is an array of inductively coupled split-ring resonators, where a linear/nonlinear impurity ring is obtained by a change in its resonance frequency by inserting a linear/nonlinear dielectric inside its slit. In the absence of the impurity, the modes are  magneto-inductive plane waves, and when a capacitive impurity is introduced, a localized mode is created.  The nonlinear impurity concept has also been found useful in studies of embedded solitons\cite{malomed}.

Now, a usual approach when dealing with nonlinear impurities is to make an educated guess about the shape of the mode (usually exponential) which then leads to the mode energy and exact spatial profile. However, this procedure might work only partially. For one thing, in the presence of nonlinearity, the number of modes depends on the available energy content, and there are possible bifurcation separating different modes with different stabilities.  Also, for problems involving boundaries, like impurities close to a surface, the need for a more serious treatment is needed.
An elegant method for dealing with impurity problems is the technique of lattice Green functions\cite{green, barton,duffy}. Originally devised for linear problems, it has been shown that it can also be extended to simple nonlinear problems\cite{molina2,molina3,molina4,molina5}.  This is the method we will follow in this work, with the added feature of fractionality.

Fractionality is a concept that has gained considerable interest in recent years. It all started a long time back from a letter exchange between Leibnitz and L'Hopital about possible generalizations of the concept of a derivate and whether it made sense to ask questions such as: what is the half derivate of a function? 
The starting point was the calculation of $d^{\alpha} x^{k}/ dx^{\alpha}$, for  $\alpha$ a real number. This means
\be 
{d^{n} x^k\over{d x^n}}= {\Gamma(k+1)\over{\Gamma(k-n+1)}} x^{k-n} \rightarrow {d^\alpha x^k\over{d x^\alpha}} = {\Gamma(k+1)\over{\Gamma(k-\alpha+1)}} x^{k-\alpha}.\label{eqx}
\ee
where $\Gamma(x)$ is the Gamma function. From Eq.(\ref{eqx}) the fractional derivative of an analytic function $f(x)=\sum_{k} a_{k} x^{k}$ can be computed by deriving the series  term by term. However, this basic procedure is not exempt from ambiguities. For instance, $(d^\alpha/d x^{\alpha})\ 1=(d^\alpha x^{0}/d x^\alpha)=(1/\Gamma(1-\alpha)) x^{-\alpha}\neq 0$, according to Eq.(\ref{eqx}). However, one could have also taken $(d^{\alpha-1}/d x^{\alpha-1})(d/dx)\ 1=0$. The initial studies were followed later by rigorous work by several mathematicians including Euler, Laplace, Riemann, and Caputo, to name some and  promoted fractional calculus from a mathematical curiosity to a full-blown research field\cite{fractional1,fractional2,fractional3,hilfer}. Several possible definitions for the fractional derivative have been given, each one with its own advantages and disadvantages. The Riemann-Liouville form is one of the most commonly used definitions, and is given by
\be
\left({d^{\alpha}\over{d x^{\alpha}}}\right) f(x) = {1\over{\Gamma(1-\alpha)}} {d\over{d x}} \int_{0}^{x} {f(s)\over{(x-s)^{\alpha}}} ds
\ee
another common form, is the Caputo formula
\be
\left({d^{\alpha}\over{d x^{\alpha}}}\right) f(x) = {1\over{\Gamma(1-\alpha)}} \int_{0}^{x} {f'(s)\over{(x-s)^{\alpha}}} ds
\ee
where, $0<\alpha<1$. This formalism that extends the usual integer calculus to a fractional one, with its definitions of a fractional integral and fractional 
derivative, has found application in several fields: fluid mechanics\cite{fluid2}, fractional kinetics and anomalous diffusion\cite{metzler,sokolov,zaslavsky}, strange kinetics\cite{shlesinger}, fractional quantum mechanics\cite{laskin,laskin2}, Levy processes in quantum mechanics\cite{levy}, plasmas\cite{plasmas}, electrical propagation in cardiac tissue\cite{cardiac}, biological invasions\cite{invasion}, and epidemics\cite{epidemics}.\\

\noindent
{\bf 2.\ The model}.\\

Let us consider a general excitation propagating along a one-dimensional, periodic lattice that contains a nonlinear impurity. The evolution equations are 
\be
i\ {d C_{n}\over{d t}} + V (C_{n+1} + C_{n-1}) + \delta_{n,0}\,\chi |C_{n}|^\beta C_{n}=0\label{eq1}
\ee
where $C_{n}$ is the amplitude for finding the excitation at site $n$ at time $t$, $V$ is the coupling coefficient, $\chi$ is the nonlinear parameter and $\beta$ is the nonlinearity exponent. In this work we will restrict ourselves to the most common exponents, $\beta=0\ \mbox{and}\ 2$ cases (linear and Kerr cases). The kinetic energy term in Eq.(\ref{eq1})
$V(C_{n+1} + C_{n-1})$, is essentially a discrete laplacian 
$\Delta_{n} C_{n}=C_{n+1}-2 C_{n}+C_{n-1}$, so that Eq.(\ref{eq1}) can be cast as
\be
i {d C_{n}\over{d t}} + 2 V C_{n} + V \Delta_{n} C_{n} + \delta_{n,0}\,\chi |C_{n}|^\beta C_{n}=0.\label{eq2}
\ee

We proceed now to replace the discrete Laplacian $\Delta_{n}$ by its fractional form 
 $(\Delta_{n})^\alpha$ in Eq.(\ref{eq2}). The form of this fractional discrete laplacian is given explicitly by\cite{discrete laplacian}:
\be
(-\Delta_{n})^\alpha C_{n}=\sum_{m\neq n} K^\alpha(n-m) (C_{n}-C_{m}),\hspace{0.5cm}0<\alpha<1 \label{delta}
\ee
where,
\be
K(m) = C_{\alpha}\  {\Gamma(|m|-\alpha)\over{\Gamma(|m|+1+\alpha)}}.\label{eq7}
\ee
and,
\be
C_{s}={4^s \Gamma(\alpha+(1/2))\over{\sqrt{\pi}|\Gamma(-\alpha)|}},\label{eq8}
\ee
where $\alpha$ is the fractional exponent. 
After replacing Eq.(\ref{delta}) into Eq.(\ref{eq2}) and searching for a stationary-state mode $C_{n}(t) = \exp(i \lambda t) \phi_{n}$, we obtain the following system of nonlinear difference equations for $\phi_{n}$:
\be
(-\lambda + 2 V )\ \phi_{n} + V \sum_{m\neq n} K^\alpha(n-m) (\phi_{m}-\phi_{n}) + \delta_{n,0}\,\chi\ \phi_{n}^{\beta+1}=0\label{eq9}
\ee
where, without loss of generality $\phi_{n}$ can be chosen as real. As can be seen from Eq.(\ref{eq9}), the presence of fractionality introduces nonlocal interactions via the symmetric kernel $K^\alpha(n-m)$. Using the relation $\Gamma(n+\alpha)\rightarrow \Gamma(n)\,n^{\alpha}$ at large $n$, we obtain the asymptotic expression $K^\alpha (m)\rightarrow 1/|m|^{1+2 \alpha}$

In the absence of any impurity, $\chi=0$, we have solutions in the form of plane waves: $\phi_{n}=A\, \exp(i k n)$. After inserting this form into Eq.(\ref{eq9}) we obtain after some simple algebra the dispersion relation
\begin{widetext}
\be
\lambda(k) = 2V - 4 V \sum_{m=1}^{\infty} K^\alpha(m) \sin((1/2) m k)^2\label{dispersion}
\ee
or, in closed form
\be
\lambda(k)=2 V - {16 V\ \Gamma(\alpha+(1/2))\over{\sqrt{\pi}\ \Gamma(1+\alpha)}}\Big( 1-\exp(-i k)\ \alpha\ \Gamma(1+\alpha)[\ R(1,1-\alpha,2+\alpha;\exp(-i k))+\exp(2 i k) \ R(1,1-\alpha,2+\alpha;\exp(i k))\ ] \Big)\label{dispersion2}
\ee
\end{widetext}
where $R(a,b,c;z)={}_2 F _{1}(a,b,c;z)/\Gamma(c)$ is the regularized hypergeometric function.  Inspection of Eq.(\ref{dispersion2}) reveals that the  bandwitdth $\Delta \lambda = \lambda(0)-\lambda(\pm \pi)$ changes with $\alpha$, increasing from a minimum value of $V$ (at $\alpha=0$) up to $4 V$ (at $\alpha=1$). As $\alpha$ decreases, the range of $K(m)$ increases causing an increase in the effective range of the coupling among sites. In the limit $\alpha\rightarrow 0$, all sites become similarly coupled, and the resulting system is similar to what is known in the literature as a simplex\cite{simplex1,simplex2}. See more on this below.

As is well-known, in tight-binding systems such as (\ref{eq9}), the evolution equations can be derived from a Hamiltonian. The Hamiltonian formalism is useful when employing Green functions to compute the behavior of the fractional system with an impurity. In our case, the Hamiltonian can be written as
\be
H = H_{0} + H_{1}
\ee
where,
\begin{eqnarray}
H_{0} &=& \sum_{n} \epsilon_{n}|n\rangle\langle n|\, + \sum_{n,m} |m\rangle V_{n,m}\,\langle n|\\
 H_{1} &=&  \chi\, |\phi_{0}|^\beta|0\rangle\langle 0|\label{H1}
\end{eqnarray}
with
\be
\epsilon_{n} = 2 V - V \sum_{m\neq n} K^\alpha(n-m)
\ee
and
\be
V_{n m}=K^\alpha(n-m)
\ee
and we have used Dirac's notation. The Hamiltonian $H_{0}$ is the `unperturbed' Hamiltonian in the absence of the impurity, while $H_{1}$ is the perturbation due to the presence of the impurity. The equations of motion are given by $i\,d C_{n}/d t = \partial H/\partial C_{n}^{*}$.

The Green function is defined as
\be
G(z) = {1\over{z-H}}\label{G}
\ee
while the unperturbed Green function is given by $G^{(0)}=1/(z-H_{0})$. In an explicit form it can be written as
\be
G_{n m}^{(0)}(z)={1\over{2 \pi}} \int_{-\pi}^{\pi} {e^{i k (n-m)} dk\over{z-\lambda(k)}}\label{G0}
\ee
where $n$ and $m$ are lattice positions, and $\lambda(k)$ is given by Eq.(\ref{dispersion2}), and where we have used the notation $G_{m n}^{(0)} = \langle m|G^{(0)}|n \rangle$. Treating $H_{1}$ as a perturbation, we can expand $G(z)$ as
\be
G(z) = G^{(0)} + G^{(0)}\ H_{1}\  G^{(0)} + G^{(0)}\ H_{1}\ G^{(0)}\ H_{1}\  G^{(0)} \cdots
\ee
After inserting (\ref{H1}) for $H_{1}$, and after resuming the perturbative series to all orders, we obtain
\be
G(z) = G^{(0)} + \chi {G^{(0)}|0\rangle \ |\phi_{0}|^\beta\  \langle 0| G^{(0)} \over{1 - \chi |\phi_{0} |^\beta \ G_{0 0}^{(0)}}}
\ee
 According to the general theory\cite{economou1}, the energy $z_{b}$ of the bound state is given by the poles of $G_{0 0}(z)$ 
\be
1 = \chi |\phi_{0}|^\beta G_{0 0}^{(0)}(z_{b}),
\ee 
while the square of the mode amplitude at site $n$ is given by the residue of $G_{n m}(z)$ at the pole
\be
|\phi_{n}|^2 = -{{G_{n 0}^{(0)}}^2 (z_{b})\over{G_{0 0}'^{(0)}(z_{b})}}.
\label{phin2}
\ee
In particular, the amplitude at the impurity site $|\phi_{0}|^2$ is given by $-G_{0 0}^2(z_{b})/G_{0 0}'^{(0)}(z_{b})$. Inserting this back into the equation for the bound state energy, we obtain a nonlinear equation for $z_{b}$:
\be
{1\over{\chi}} = {G_{0 0}^{(0){\beta+1}}(z_{b})\over{[-G_{0 0}'^{(0)}(z_{b})]^{\beta/2}}}.\label{zb}
\ee
 
The Green function formalism is also useful to compute the transmission of plane waves across the impurity. From the general formalism\cite{green}, the transmission amplitude $T\sim 1/|1-\epsilon G_{0 0}^{+}(z)|^2$, while the reflection amplitude $R\sim \epsilon^2 |G_{0,0}^{+}(z)|^2/|1-\epsilon G_{0 0}^{+}(z)|^2$, where, in our case $\epsilon=\chi\,|\phi_{0}|^\beta$. Let us define a normalization factor $N= T + R$; then the transmission coefficient $t=T/N$ and the reflection coefficient $r=R/N$ will be given by
\be
t(z) = {1\over{1 + \epsilon^2 |G_{0 0}^{+}(z)|^2}}\label{t}
\ee
\be
r(z)={\epsilon^2 |G_{0 0}^{+}(z)|^2\over{1 + \epsilon^2 |G_{0 0}^{+}(z)|^2}}, \label{r}
\ee
\begin{figure}[t]
 \includegraphics[scale=0.25]{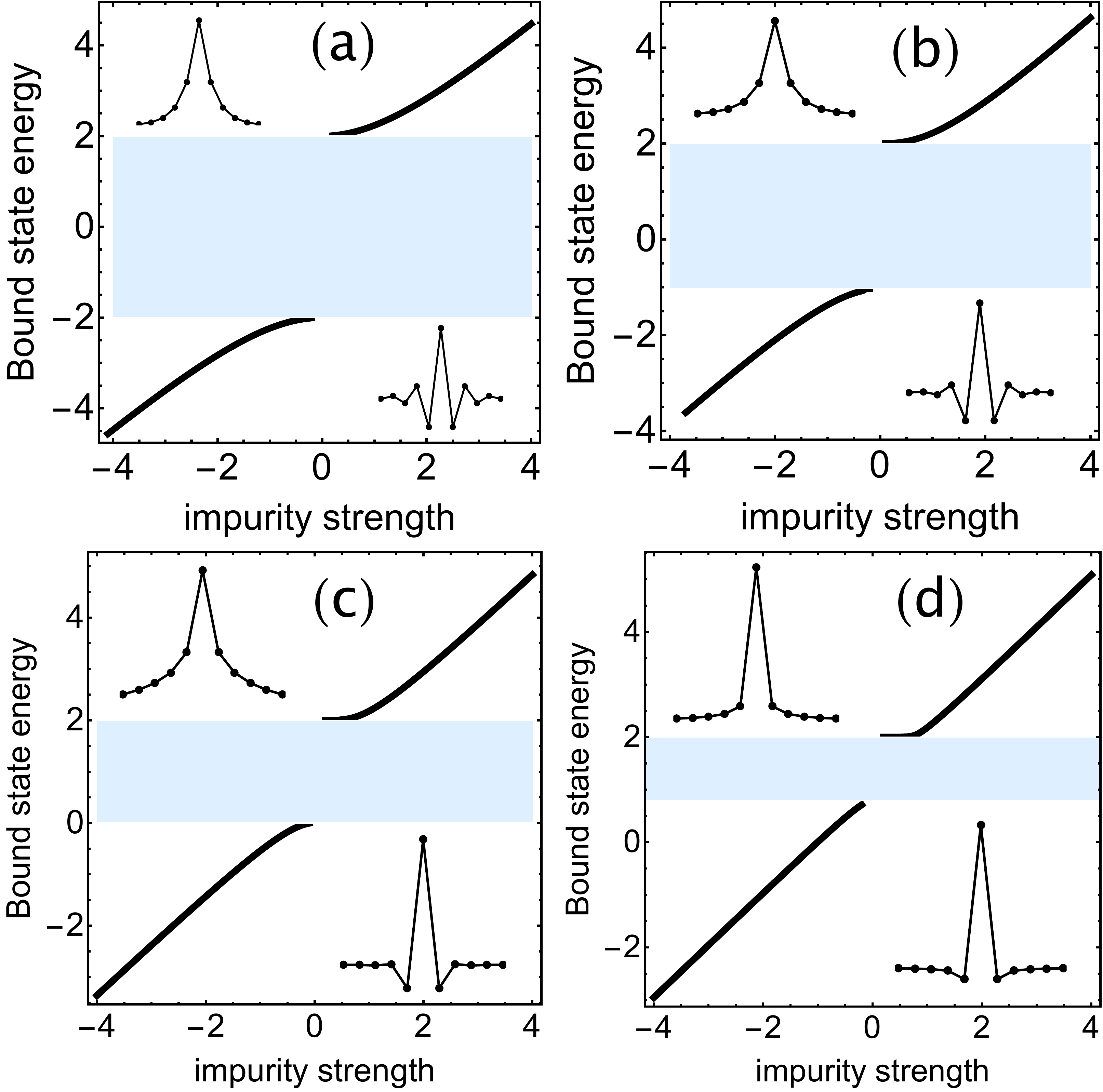}
  \caption{Linear impurity ($\beta=0$): Bound state energy vs impurity strength for several fractional exponents: (a) $\alpha=1$, (b) $\alpha=0.8$ (c) $\alpha=0.5$ (d) $\alpha=0.2$. Some representative impurity profiles are also shown.}  \label{fig1}
\end{figure}
\begin{figure}[t]
 \includegraphics[scale=0.25]{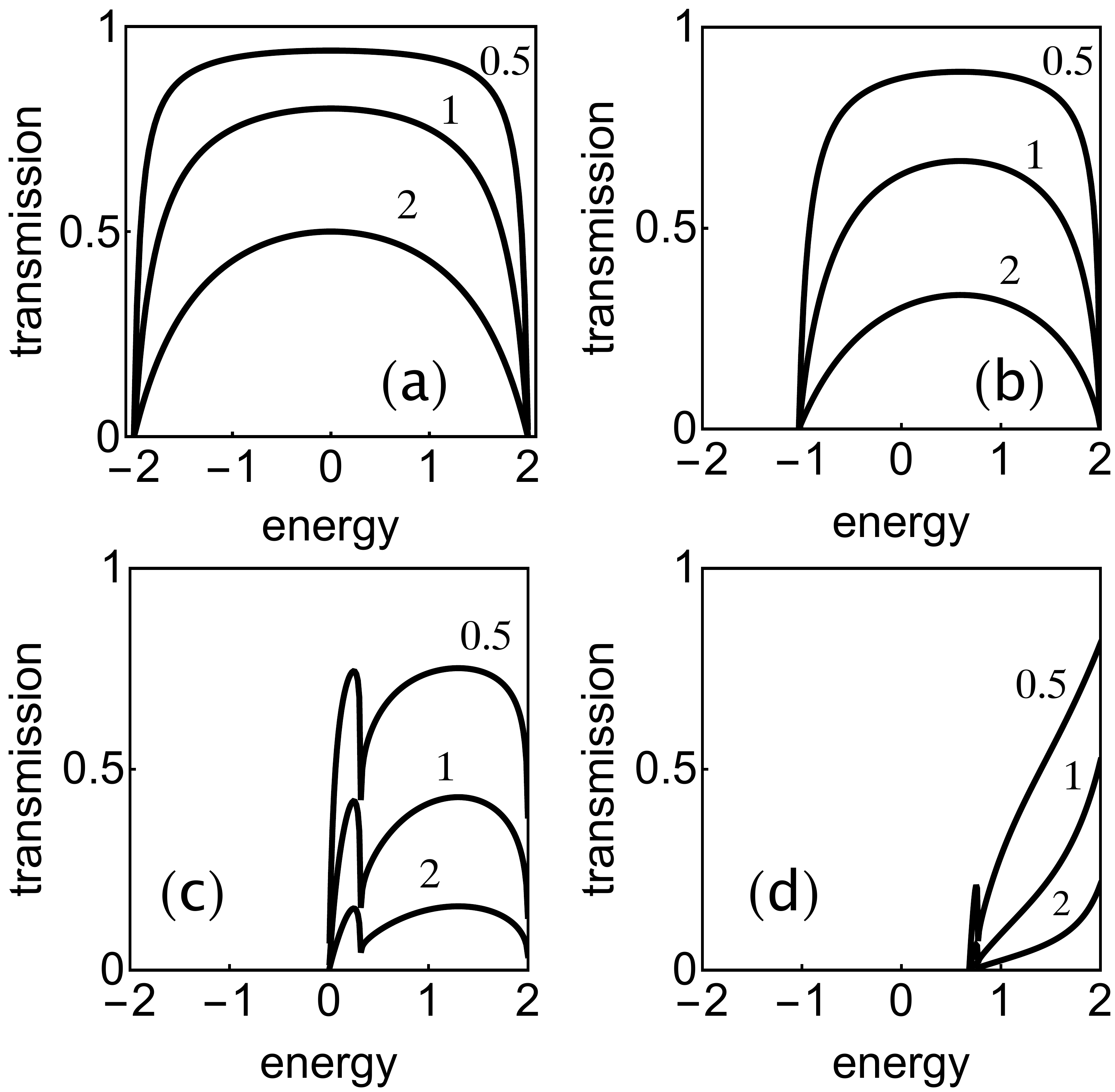}
  \caption{Transmission coefficient for plane waves across a linear impurity, for several fractional exponents. (a) $\alpha=1$, (b) $\alpha=0.8$, (c) $\alpha=0.5$, (d) $\alpha=0.2$. The numbers on each curve denote the value of the impurity strength $\chi$.}  \label{fig1}
\end{figure}
\begin{figure}[t]
 \includegraphics[scale=0.23]{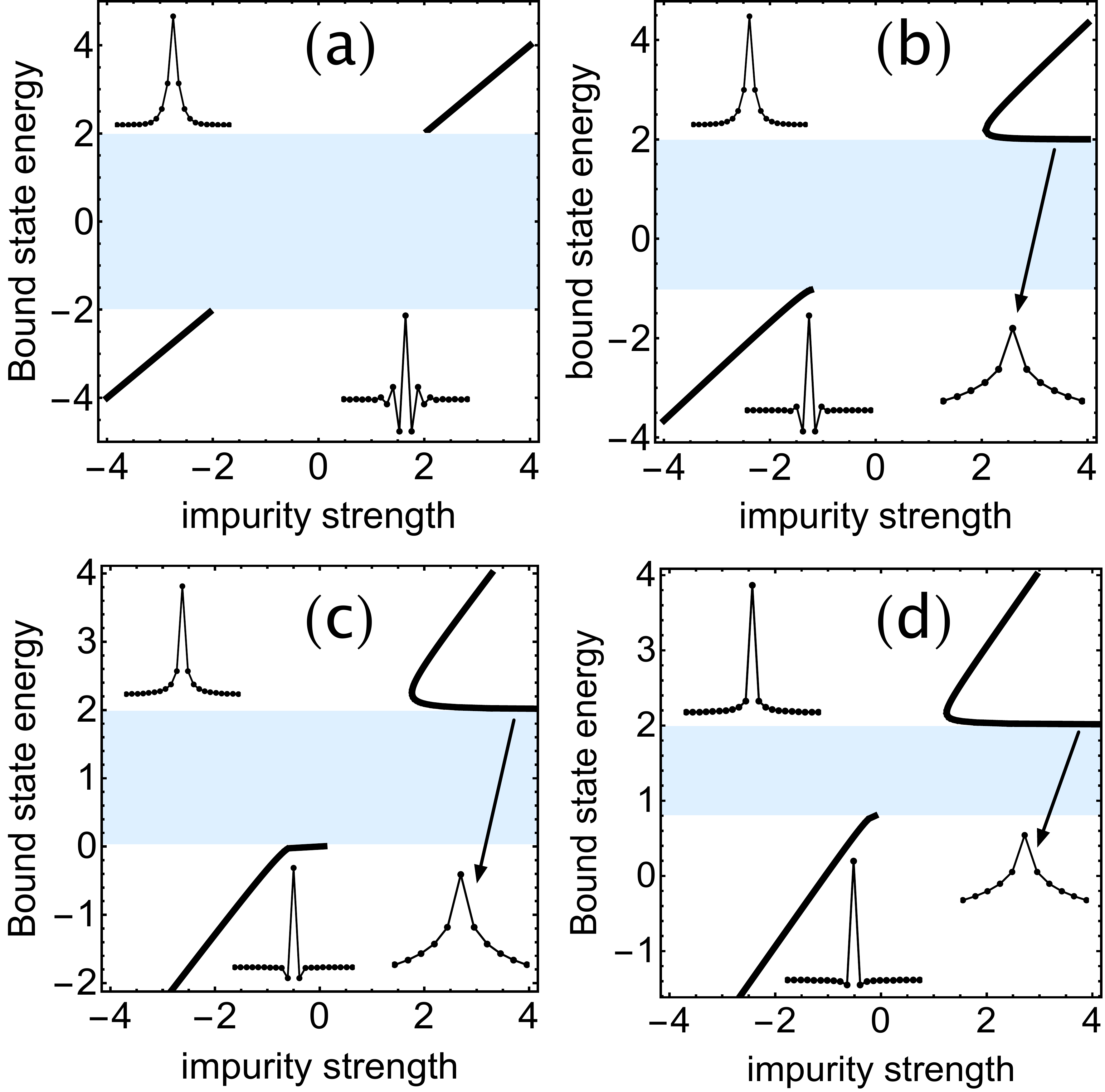}
  \caption{Nonlinear impurity ($\beta=2$): Bound state energy vs impurity strength for several fractional exponents: (a) $\alpha=1$, (b) $\alpha=0.8$ (c) $\alpha=0.5$ (d) $\alpha=0.2$. Typical representative profiles are also shown.}  \label{fig1}
\end{figure}
where $G^{+}(z) = \lim_{\eta\rightarrow 0} \,G(z+i \eta)$ and $z$ is inside the band $\lambda(k)$ given by (\ref{dispersion2}). Clearly, $t + r =1$. Also, given that the transmission coefficient is also given by the probability at the impurity site, $t=|\phi_{0}|^2$, this means $\epsilon = \chi\, t^{\beta/2}(z)$. Then, Eq.(\ref{t}) becomes a nonlinear equation for $t(z)$, except in the linear case ($\beta=0$) where it reduces to a closed form for $t(z)$. In our case, we are interested in the Kerr nonlinearity $\beta=2$, where Eq.(\ref{t}) becomes a cubic equation:
\be
b\,t^3+t-1=0\hspace{0.5cm} \mbox{with}\ b = \chi^2\, |G^{+}(z)|^2.
\ee
with real solution
\be
t = {(9 b^2+\sqrt{3}\sqrt{4 b^3+ 27 b^4})^{1/3}\over{2^{1/3}3^{2/3} b}}-{(2/3)^{1/3}\over{(9 b^2+\sqrt{3}\sqrt{4 b^3+ 27 b^4})^{1/3}}}
\ee
Where is the fractionality of the system? It is hidden in $b$ through $G^{+}(z)$ which depends 
on $\lambda(k)$, which in turn depends directly on $\alpha$, as shown by Eq.(\ref{dispersion2}).
\\

\noindent
{3.\ \bf Results}. \\

With the help of Eqs.(\ref{G0}),(\ref{phin2}), (\ref{zb}) and (\ref{t}) we are ready to compute numerically the bound state energy $z_{b}$ and profile $|\phi_{n}|^2$, as well as the transmission across the impurity for different values of the fractional exponent $\alpha$. The fractionality is hidden inside $\lambda(k)$ and its complexity preclude us to 
obtain $G_{n m}^{(0)}(z)$ in closed form.

\noindent
{\em (a) Linear case ($\beta=0$)}.\\
Figure 1 shows results for the  bound state energy of the linear impurity, as a function of $\chi$, the impurity strength. We see that as soon $\chi$ is different from zero, there is a bound state with energy above the band. We notice that, as the fractional exponent decreases, the bandwidth decreases steadily. The curves of $z_{b}$ vs $\chi$ seem to converge to a straight line at large impurity strength values. This is proved as follows: From Eq.(\ref{G0}), we see that at large $z$, $G_{0 0}^{(0)}(z)\rightarrow 1/z$. After inserting this into (\ref{zb}), we obtain $z_{b}\rightarrow \chi$. This is valid for any $\beta$.
\begin{figure}[t]
 \includegraphics[scale=0.21]{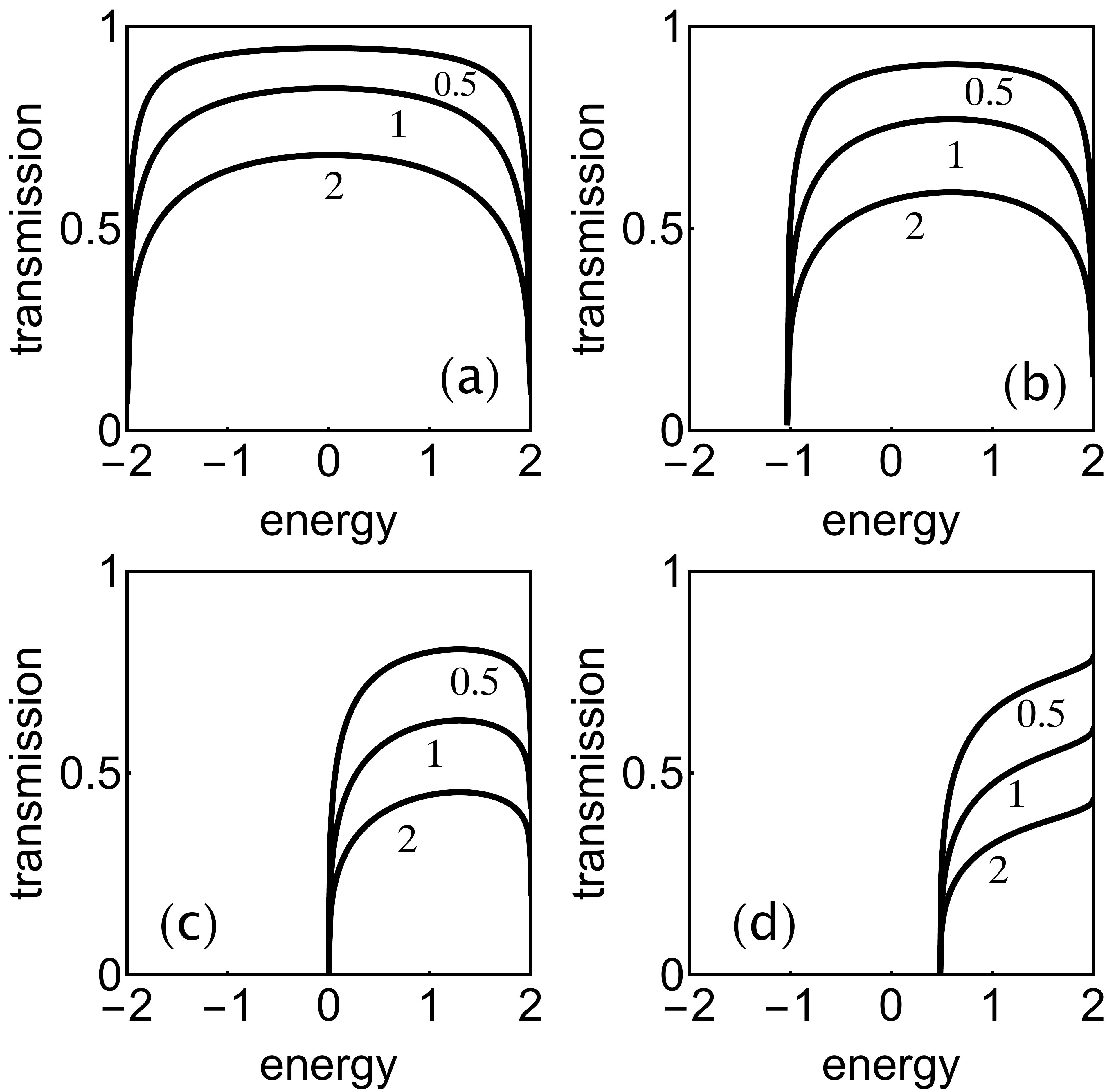}
  \caption{Transmission coefficient for plane waves across a nonlinear impurity, for several fractional exponents. (a) $\alpha=1$, (b) $\alpha=0.8$, (c) $\alpha=0.5$, (d) $\alpha=0.2$. The numbers on each curve denote the value of the impurity strength $\chi$.}  \label{fig1}
\end{figure}
\begin{figure}[t]
 \includegraphics[scale=0.21]{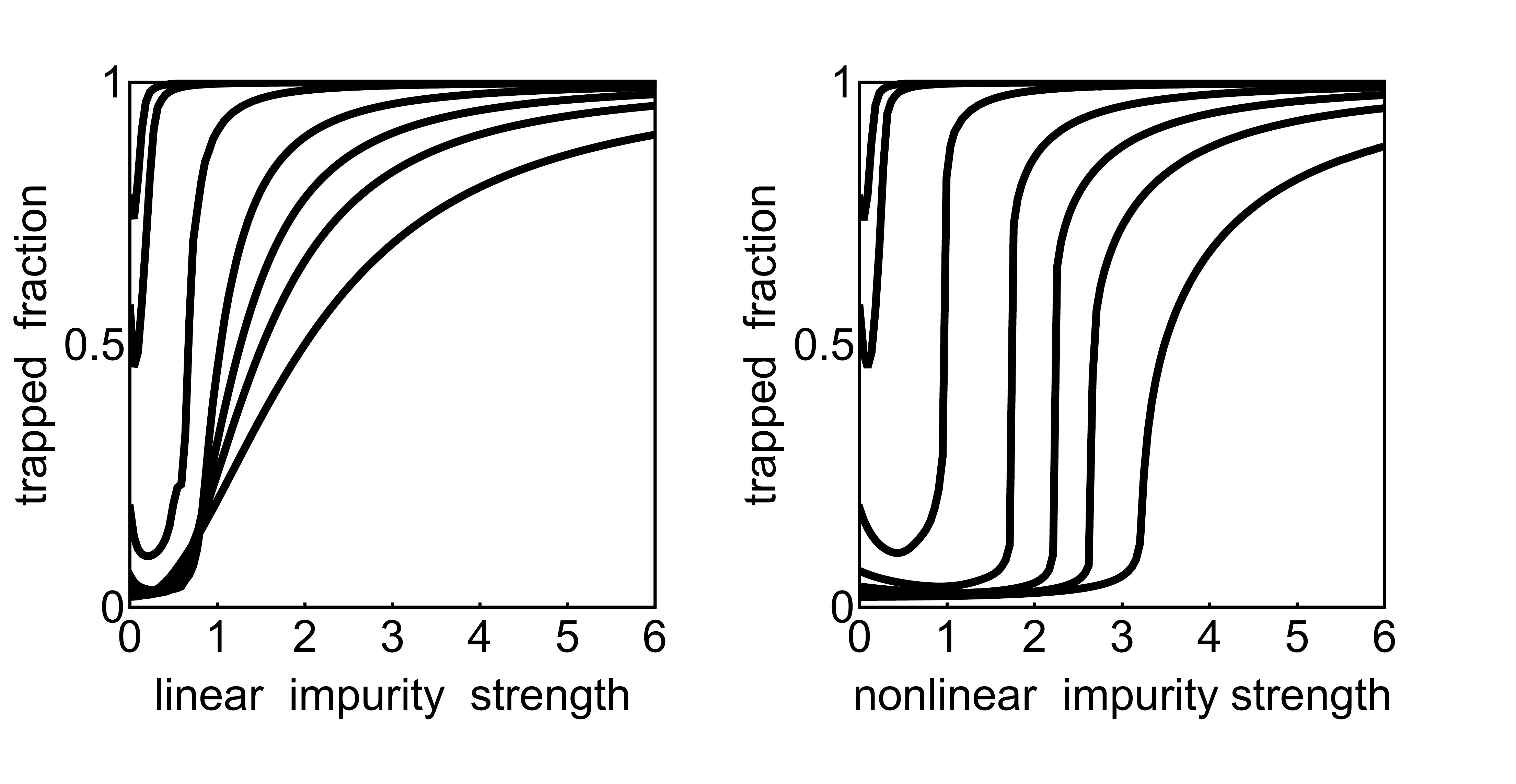}
  \caption{Time-averaged trapped fraction at impurity site versus nonlinearity strength, for the linear (left) and nonlinear (right) cases and for several fractional exponent values. From left to right: $\alpha=0.01, 0.02, 0.1, 0.3, 0.5, 0.7, 1.0$.}  \label{fig1}
\end{figure}
Also, we observe that, as $\alpha$ is decreased, the energy curves seem to reach the asymptotic regime of large impurity sooner. Figure 1 also shows some examples of the bound state profiles. We also note that, for $\alpha=1$, i.e., the standard case, the modes divide themselves into two classes: an upper branch with an unstaggered profile, for positive impurity strength and positive energy, and a lower branch with a staggered profile for negative impurity strength and negative energy. In other words, the mode obeys the  staggered-unstaggered symmetry
 $\{\phi_{n}, z_{b}, \chi \}\leftrightarrow \{ (-1)^n \phi_{n}, -z_{b}, -\chi\}$. 
Now, as $\alpha$ decreases from unity, the staggered-unstaggered character is progressively lost due to the extended character of the coupling at small $\alpha$ which destroys the symmetry.
 
Figure 2 shows the transmission coefficient of a plane wave across a linear impurity, for several fractional exponents $\alpha$. As mentioned above, as $\alpha$ decreases the bandwidth decreases, and the transmission curve gets more and more compressed into a narrow energy region. This implies an enhanced tendency towards degeneration. Also, for a given $\alpha$, an increase in $\beta$ decreases the transmission, a feature that is expected on general grounds. For small $\alpha$ values, $t(z)$ acquires an extra maximum, near the lower band edge. Finally at even smaller $\alpha$, $t(z)$ becomes monotonically increasing with $z$. These are the interesting cases since at small $\alpha$ the range of the coupling exceeds one, and the system becomes long-range. In a long-range system is hard to set up a transmission problem in the usual direct way since the scattering region has no clear boundaries. Thus, the usefulness of the Green function approach.
 
\noindent
{\em (b) Nonlinear case ($\beta=2$)}.\\
Figure 3 shows results for the  bound state energy of the nonlinear impurity, as a function of $\chi$, the impurity strength. Unlike the linear impurity case, a minimum impurity strength is needed to create a bound state. Also, a second bound state appears in the upper branch. 
The energy of this extra state approaches the upper band edge ($z=2$), as the impurity increases. The energy of the other mode in the upper branch behaves as in the linear case, i.e., $z_{b}\rightarrow \chi$ at large $\chi$. The lower branch features only a single mode whose energy approaches $-|\chi|$ at large $|\chi|$ values. The impurity mode profiles are similar to the ones encountered in the linear case. The staggered-unstaggered symmetry is not obeyed, except at $\alpha=1$. 

Figure 4 shows the transmission coefficient of plane wave across the nonlinear impurity, for several fractional exponents $\alpha$. The transmission plots look similar to the linear ones, at first glance: We see that same band narrowing of the band with decreasing $\alpha$, also the decrease in transmission as the impurity strength is increased. We notice, however that the transmission curves are flatter than in the linear case. We ascribe this to nonlinearity since it was observed first for the standard ($\alpha=1$) nonlinear impurity.  Also, the nonlinear transmission is always devoid of intermediate maxima. At small $\alpha$ no secondary maxima are present.

Finally, we investigate the influence of fractionality on the well-known phenomenon of the selftrapping transition. It can be observed when an impurity is placed on a single site initially. The time evolution shows that a critical threshold of nonlinearity strength exists below which the excitation diffuses away in a ballistic manner, while above threshold, a finite portion of the excitation remains at the initial site. The trapped portion increases with nonlinearity.
To display the selftrapping transition we do a plot of the time-average probability at the initial site vs the nonlinearity strength. We do this for different fractional exponents. Figure 5 shows the results for the linear (left side) and for the nonlinear (right side) case. In both graphs we plot the (time-averaged) trapped fraction vs the impurity strength. For both cases, we see that a decrease in $\alpha$ increases the amount of trapping, but while in the linear case no trapping transition is observed, for the nonlinear case a clear transition is observed, with the critical nonlinearity threshold shifting to lower nonlinearity values as the fractional exponent decreases. Now, in both cases we do observe that, at the smallest fractional exponent $\alpha$ values, the amount of trapping increases towards unity at zero impurity strength. We believe this phenomenon is directly related to the long-range coupling $K^{\alpha}(n)$ that is manifest at small fractional exponents, where a given site is effectively coupled to sites that are far away. As mentioned before, $K^\alpha (n)\rightarrow 1/|n|^{1+2 \alpha}$, meaning that at $\alpha\sim 0$, the coupling decreases as the inverse of the distance only, in marked difference with standard, 
non-fractional  tight-binding model where the coupling decreases exponentially with distance. In the limit of zero exponent, all sites are approximately equally coupled, and that makes the system mathematically equivalent to a {\em simplex}\cite{simplex2,simplex1}. In a simplex all sites of a cluster interact equally with all the others. If an excitation is placed on a single site, the probability for finding the excitation on the initial site some time later is given by\cite{simplex1}
\be
|\phi_{0}(t)|^2 = {(N-1)^2 + 1\over{N^2}} + 2 \left( {N-1\over{N^2}}\right)\, \cos(N t)
\ee
whose time average gives
\be
\langle |\phi_{0}(t)|^2 \rangle = {(N-1)^2 + 1\over{N^2}}
\ee
where $N$ is the number of sites, Thus, for a large number of sites, the trapped fraction tends to unity. This is clearly shown in Fig.5.
\\

{\bf 4.\ Conclusions}.\\

In this work, we have examined the consequences of using a fractional form of the discrete Laplacian, on the stationary and dynamical properties of a linear/nonlinear impurity embedded in a one-dimensional lattice. The degree of fractionality depends on a single parameter, $\alpha$ ($0<\alpha<1$) whose departure from unity induces an effective nonlocal coupling among sites. In the absence of the impurity, we obtained the band structure in closed form, showing that a decrease in $\alpha$ induces a narrowing of the bandwidth. In fact, for $\alpha\rightarrow 0$, we have a near-degeneracy of the spectrum where all energies collapse to the value $\lambda(k)=V$ (except for the single wavevector $k=0$, where $\lambda(0)=2 V$). In the presence of a single impurity, we decided to resort to the formalism of lattice Green functions, since it is a powerful method that gives rigorous results, especially in the nonlinear domain where it might not be clear at first the number of possible bound states, or critical parameter values separating different physical behaviors. When the impurity is present, the shape of the localized mode profiles remains like those of the non-fractional case. The long-range effective coupling leads to a loss of the staggered-unstaggered symmetry that is present at $\alpha=1$. Yet, the modes still show traces of this symmetry, especially at $\alpha$ values not far from unity. For the linear case, the formalism predicts that only one state is possible at any impurity strength, for any fractional exponent. For the nonlinear impurity case, however, the formalism predicts that up to two bound states are possible for impurity strengths beyond a certain threshold, which depends on the value of $\alpha$.

The transmission of plane waves across the impurity was also computed with the help of Green functions. That's because an $\alpha<1$ induces a long-range coupling that makes the impurity be coupled over many sites, making it difficult to set up the transmission problem in a straightforward manner. The most salient feature we observe is that for both, linear and nonlinear impurity, a decrease of fractional exponent decreases the overall transmission. For the linear case and for mid-values of $\alpha$, the transmission develops a second maximum. For the smallest $\alpha$, the transmission becomes monotonically increasing  with the energy. 

The trapping of an excitation on the impurity site shows that in general, it increases with decreasing $\alpha$. In particular, in the nonlinear case, the well-known trapping transition for $\alpha=1$ gets shifted to lower impurity strength values as $\alpha$ is decreased. At the same time, the trapping at small impurity strength values also increases with decreasing $\alpha$, a behavior that is ultimately due to the near-degeneracy condition at the smallest $\alpha$ values.

{\bf Acknowledgments}

This work was supported by Fondecyt Grant 1200120.

\end{document}